\documentclass[aps,prd,twocolumn,showpacs,notitlepage,superscriptaddress,nofootinbib,showkeys]{revtex4-1}
\pdfoutput=1
\usepackage{graphicx}
\usepackage{amsmath}
\usepackage{amsfonts}
\usepackage{amssymb}
\usepackage{xcolor, soul}
\usepackage{epstopdf}
\usepackage{lipsum}
\usepackage{float}
\usepackage{subfigure}
\usepackage{hyperref}
\hypersetup{
     colorlinks   = true,
     citecolor    = red,
     linkcolor    = blue,
     urlcolor     = cyan,
}
\def\beq{\begin{equation}}
\def\eeq{\end{equation}}
\def\bea{\begin{eqnarray}}
\def\eea{\end{eqnarray}}
\def\be{\begin{equation}}
\def\ee{\end{equation}}
\def\pr{\partial}
\def\nno{\nonumber}
\def\bse{\begin{subequations}}
\def\ese{\end{subequations}}

\def\tk{\tilde{k}}
\def\tg{\tilde{g}}
\def\tu{\tilde{u}}
\def\bk{\textbf{k}}
\def\bx{\textbf{x}}

\graphicspath{{./figs/}}
\begin{document}

\title{Magnetogenesis from Anisotropic Universe 
}
\author{Sourav Pal}%
\email{pal.sourav@iitg.ac.in}
\affiliation{%
	Department of Physics, Indian Institute of Technology Guwahati,\\
	Guwahati 781039, Assam, India 
}%
\author{Debaprasad Maity}
\email{debu@iitg.ac.in}
\affiliation{%
	Department of Physics, Indian Institute of Technology Guwahati,\\
	Guwahati 781039, Assam, India 
}%

\author{Tuan Q. Do }%
\email{tuan.doquoc@phenikaa-uni.edu.vn}
\affiliation{Phenikaa Institute for Advanced Study, Phenikaa University,  Hanoi 12116,  Vietnam}
\affiliation{ Faculty of Basic Sciences, Phenikaa University, Hanoi 12116, Vietnam}

\date{\today}

\begin{abstract}
The existence of large-scale anisotropy can not be ruled out by the cosmic microwave background (CMB) radiation. Over the years, several models have been proposed in the context of anisotropic inflation to account for CMB's cold spot and hemispheric asymmetry. However, any small-scale anisotropy, if exists during inflation, is not constrained due to its nonlinear evolution in the subsequent phase. This small-scale anisotropy during inflation can play a non-trivial role in giving rise to the cosmic magnetic field, which is the subject of our present study. Assuming a particular phenomenological form of an anisotropic inflationary universe, we have shown that it can generate a large-scale magnetic field at $1$-Mpc scale with a magnitude $\sim 4\times 10^{-20}~G$, within the observed bound. 
Because of the anisotropy, the conformal flatness property is lost, and the Maxwell field is generated even without explicit coupling. This immediately resolves the strong coupling problem in the standard magnetogenesis scenario. 
In addition, assuming very low conductivity during the reheating era, we can further observe the evolution of the electromagnetic field with the equation of state (EoS) $\omega_{eff}$ and its effects on the present-day magnetic field.
\end{abstract}
\maketitle
\section{Introduction} \label{intro}

It is well known that our universe is magnetized on all observational scales, from planets and stars to large-scale galaxies and galaxy clusters. In particular, the magnetic field strength has been observed in the range from $\mu G$ for galaxies and galaxy clusters to a few $G$ for planets and $10^{12}$ $G$ for neutron stars. From Gamma-ray observations and Faraday rotation measurements, the magnetic field in the intergalactic medium (IGM) has also been shown to be bounded with the strength ranging from $10^{-10}-10^{-22}$ $G$ \cite{Grasso:2000wj, Beck:2000dc, Widrow:2002ud, Kandus:2010nw, Durrer:2013pga}. It is possible that the primordial magnetic fields on a large scale ($\sim 1~ MPc$) were generated during the Big Bang or later and survived until today as a relic. The origin of the magnetic field in galaxies and galaxy clusters can be explained through classical magneto-hydrodynamic processes magnifying the tiny seed magnetic field. It is important to identify the origin of the primordial magnetic fields. There have been some proposed mechanisms for generating large-scale primordial magnetic fields, which can be found in the interesting review papers listed in Refs. \cite{Subramanian:2015lua, Kulsrud:2007an, Brandenburg:2004jv, Subramanian:2009fu, Sharma:2017eps,Sharma:2018kgs, Jain:2012ga,Durrer:2010mq, Kanno:2009ei,Campanelli:2008kh, Demozzi:2009fu,Bamba:2012mi, Bamba:2006ga,Bamba:2003av, Bamba:2004cu,Ratra:1991bn}. Among them, the Ratra model \cite{Ratra:1991bn} is the most accepted one, where the electromagnetic fields are generated during the inflationary era by breaking the conformal invariance of the Maxwell term, i.e., $F_{\mu\nu}F^{\mu\nu}$, through non-minimal coupling(s) with other fields such as scalar fields. 

Inspired by the mechanism in \cite{Turner:1987bw, Ratra:1991bn, Dolgov:1993vg, Bamba:2006ga, Demozzi:2009fu, Martin:2007ue}, we propose another mechanism for generating the primordial magnetic fields through anisotropic spacetime. In particular, the Maxwell field experiences the existence of the anisotropy of spacetime during the inflationary era, leading to the genesis of primordial magnetic. In cosmology, there exists a nice classification of homogeneous but anisotropic spacetimes called the Bianchi universe \cite{Ellis:1968vb, Ellis:2006ba, Watanabe:2009ct, Kanno:2010nr, Do:2011zz}. It turns out that the Bianchi type I metric is the simplest one and can be regarded as a straightforward extension of the Friedmann-Lemaitre–Robertson–Walker (FLRW) spacetime. Hence, we chose the Bianchi type I metric for our study.

Remarkably, it has long been argued that the very early universe, which is close to the initial singularity, should be strongly anisotropic \cite{Belinsky:1970ew, Collins:1972tf, Belinsky:1982pk}. During an inflationary phase of the early universe, all spatial anisotropies, which could happen in a pre-inflationary phase, should decrease very quickly such that the universe speedily approaches a locally isotropic state, as pointed out in Refs. \cite{Starobinsky:1982mr,Wald:1983ky}. It should be noted that this scenario is consistent with the so-called cosmic no-hair conjecture, which claims that all initial anisotropies and inhomogeneities should disappear in a late-time universe \cite{Gibbons:1977mu, Hawking:1981fz}. Very interestingly, some recent unavoidable anomalies in the cosmic microwave background (CMB) radiations confirmed by the Planck \cite{Schwarz:2015cma, Akrami:2018odb} such as the cold spot and hemispheric asymmetry, have challenged the standard inflationary universe models, which are based on the cosmological principle stating, the universe should be homogeneous and isotropic on large scales. In addition, other interesting observational evidences, which have called the validity of cosmological principle into question, have been listed in a recent interesting review \cite{Aluri:2022hzs}. These remarkable points lead us to a possible scenario of an anisotropic inflationary universe in early times. Many papers have been working on anisotropic inflation, e.g., see Refs. \cite{Watanabe:2009ct,Kanno:2010nr,Do:2011zz} as well as Refs. \cite{Barrow:2005qv,Gumrukcuoglu:2007bx,Pitrou:2008gk,Starobinsky:2019xdp,Galeev:2021xit,Nojiri:2022idp}.

 This paper does not discuss the origin and evolution of anisotropy in spacetime. Instead, we treat it as a perturbation over the FLRW background. As the spatial anisotropy breaks the conformal flatness of the background in the electromagnetic (EM) field, we do not need any explicit coupling with the scalar field as proposed in the literature \cite{Demozzi:2009fu, Bamba:2012mi, Bamba:2006ga, Bamba:2003av, Bamba:2004cu} for gauge field production during inflation. In this context, it is important to mention the challenges in inflationary magnetogenesis, namely the strong-coupling problem and the backreaction problem \cite{Sharma:2017eps, Demozzi:2009fu, Ferreira:2013sqa, Tasinato:2014fia}. In the literature, several mechanisms and different types of coupling \cite{Ferreira:2013sqa} have been introduced to overcome these problems. However, in our formalism, the strong coupling problem is readily solved in this paper since no explicit coupling is involved with the gauge field. However, there might still be a possibility of backreaction, which we will study in detail as we proceed. 

The paper is organized as follows. (i) An introduction of the present paper has been written in Sec. \ref{intro}. (ii) In Sec. \ref{sec2}, we describe the basic formalism and the quantization of the gauge field in an anisotropic background. (iii) In Sec. \ref{sec3}, we show the evolution of the gauge field during the inflationary era and the strength of the present-day magnetic field under an instant reheating scenario. However, a scenario might occur when the universe undergoes a prolonged reheating era, affecting the magnetic field. (iv) We discuss the evolution in such a scenario in Sec. \ref {sec4}. (v) Finally,  we discuss the findings and implications of this proposal in Sec. \ref {sec5}.

\section{The setup} \label{sec2}
We introduce the spatial anisotropy in the background through the homogeneous but anisotropic Bianchi type I metric. In general, this type of metric can be written as
\beq\label{metric}
ds^2 = a^2(\eta)\left[-d\eta^2+ b^2(\eta)dx^2 +dy^2+dz^2\right],
\eeq
where $\eta$ is the conformal time, $a(\eta)$ is the overall scale factor and $b(\eta)$ is the anisotropic factor along the $x$-direction.  In our phenomenological model, we impose the condition that the anisotropy in the spacetime exists only in the inflationary era, although the spacetime remains continuous. This ansatz guarantees that the conformal flatness is restored after inflation. These conditions can be satisfied by various models of the anisotropic factor $b(\eta)$. However, we take a particular model that satisfies all the necessary conditions
\beq\label{model}
b(\eta) = 1 + \alpha~ e^{-\left(\frac{\eta}{\eta_m}\right)^2},
\eeq
In the above Eq. \eqref{model}, $\alpha$ is a dimensionless parameter that 
\begin{figure}
\begin{center}
\includegraphics[scale=0.65]{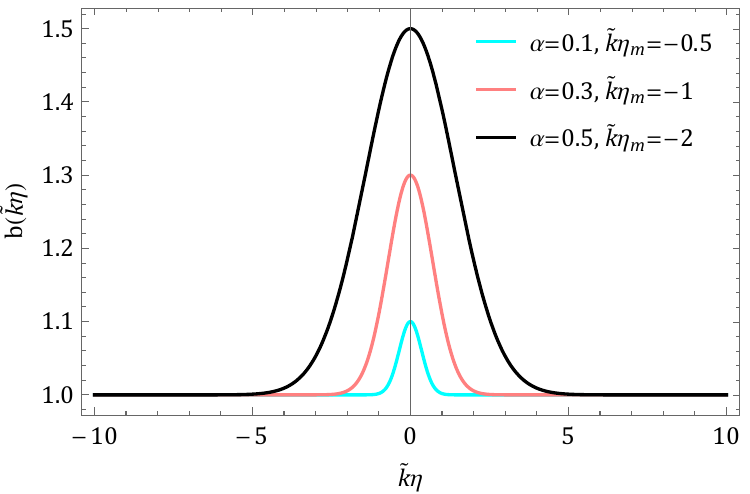}
\caption{Behaviour of the anisotropic factor $b(\tk\eta)$ with $\tk\eta$ for anisotropic free parameters  $\alpha$ and $
\tk\eta_m$.  }\label{aniso_model}
\end{center}
\end{figure}
determines the strength of the anisotropy. Furthermore, $\eta_m$ is a parameter that dictates the overall behavior of the anisotropic background. An example of the anisotropic background is shown in Fig. \ref{aniso_model}. Here, $\alpha$ and $\eta_m$ are free parameters of the anisotropic model. In this paper, we do not discuss the origin of such anisotropy. However, the anisotropy, particularly near the end of inflation, maybe a combined effect of quantum field theory and the sudden breakdown of slow-roll conditions. We will come back to this issue in the future.

The action of Einstein-scalar-vector theory can be given by
\bea
\label{action}
S= \int d^4 x \sqrt{-g}\bigg[ \frac{1}{16\pi G}R-\frac{1}{2}\partial_\mu \phi \partial^\mu \phi -V(\phi)\nno\\
-\frac{1}{4} F_{\mu\nu} F^{\mu\nu} \bigg],
\eea
where $G$ is the gravitational constant, $\phi$ is a scalar field, and $F_{\mu\nu} \equiv \partial_\mu A_\nu -\partial_\nu A_\mu$ is the field strength of the vector field $A_\mu (t,{\bf x})$ describing the electromagnetic field. 
In this paper, the dynamics of the scalar field and the metric itself due to the anisotropy present are beyond our scope. Therefore, we will mainly discuss the dynamics of the EM field during the inflationary era due to the spatially anisotropic background. Therefore, the Lagrangian of interest here is the Lagrangian corresponding to the electromagnetic field, which, according to Eq. \eqref{action}, is given by
{
\beq
{\cal L}_{em}= \frac{1}{2a^2} g^{jn} A'_j A'_n -\frac{1}{4} g^{im} g^{jn} F_{ij} F_{mn},
\eeq
where the prime denotes a derivative with respect to conformal time. In this paper, all the physical quantities are denoted with the lower index, e.g., the physical momentum is denoted by $k_i$, and the vector potential is given by $A_i$.
We set $A_0=0$ as the choice of gauge, and unlike the case of conformally flat spacetime,  the $\nu =0$ component satisfies the modified constraint  equation,
\bea
\tg^{im} \pr_i A'_m &=& 0,\label{constraint}
\eea
where we have defined $\tg^{ij}=a^2 g^{ij}$ for simplicity of calculation. It can be further shown that the above equation boils down to the usual Coulomb condition for the conformally flat case. However, the raising (and lowering) of the indices are done through the metric component $g^{ij}(g_{ij})$.
Similarly, the dynamical equation of motion for the magnetic vector potential $A_i$ can be calculated from the $\nu=j$ component, which boils down to
\bea
A''_n+ \frac{b'}{b}A'_n+ \tg_{jn}\tg^{'jk} A'_k-\tg^{im}\pr_i F_{mn}&=&0.\label{eom}
\eea
It is important to note that the metric components play a crucial role in the dynamics of the field. In the case of a standard conformally flat background, the term with the metric component's derivative vanishes, giving us the regular plane wave solutions. Now, we will promote the fields and their conjugates as operators. The conjugate momentum operator corresponding to the field operator $A_i$ turns out as
\bea\label{con_mom}
\Pi^i = \frac{\pr \mathcal{L}_{em}}{\pr A'_i}
= \frac{1}{a^2} g^{im} A'_m.
\eea

To quantize the field, we decompose the magnetic vector potential $A_i$ as,
\bea\label{mode_exp}
A_i(\eta,\bx )= \sum_p\int\frac{d^3 k}{(2\pi)^3}\bigg(a_{\bk}^{(p)}u_i^{(p)}(\eta)e^{i k_n x^n}\nno\\
+a_{\bk}^{\dagger(p)}u_i^{*(p)}(\eta)e^{-i k_n x^n}\bigg).
\eea
In the above Eq. \eqref{mode_exp}, $(p)$ is the polarization index, $a_{\bk}^{(p)}$ and $a_{\bk}^{\dagger(p)}$ are the annihilation and creation operators corresponding to the polarization mode $(p)$. They follow the general commutation relation,
\beq
\left[a_{\bk}^{(p)},a_{\bk '}^{(q)}\right]= \delta^{pq} (2\pi)^3 \delta^3(\bk-\bk ').
\eeq
In this article, the boldface letters represent vector quantities.

In this context, it is important to discuss the commutation relation of the magnetic vector potential $A_i$ and its conjugate momentum $\Pi^i$. We impose the commutation relation, such that they satisfy the constraint Eq. \eqref{constraint} on vector potential $A_i$,
\begin{eqnarray}\label{commute}
\left[A_i(\eta,\bx),A_j(\eta,{\bf y})\right] = 0 ;~ \left[\Pi_i(\eta,\bx),\Pi_j(\eta,{\bf y})\right] = 0, \\
 \left[A_i(\eta,\bx),\Pi^j(\eta,{\bf y})\right]=\frac{i}{\sqrt{-g}}\int\frac{d^3k}{(2\pi)^3} e^{i k_n(x^n-y^n)}\nno\\
\times \left(\delta_i^j-\frac{k_ik^j}{k_nk^n}\right) .
\end{eqnarray}
With the quantization of the field, we now get the mode function equations using Eq. \eqref{eom}. The mode functions satisfy the relation,
\beq\label{eom_mode_main}
u_n''+\frac{b'}{b}u_n'+ \tg^{'jl}\tg_{jn}u_l'+\tg^{im}\left(k_m k_i u_n-k_n k_i u_m\right)=0.
\eeq
The polarization index is omitted here, as all the polarization modes follow the same equation of motion.
Similarly, the constraint Eq. \eqref{constraint} in terms of the mode function becomes
\beq\label{constraint_mode}
\tg^{in} k_i u_n'= 0.
 \eeq
Interestingly, Eq. \eqref{eom_mode_main} contains the derivative of the metric coefficients, which works as the source of particle production during the inflationary era. As the mode function equation contains the derivative of the metric components, substituting the metric components, we get the modified mode function equations as,
\begin{align}
u_1'' -\frac{b'}{b}u_1'+ k_2^2 u_1 +k_3^2 u_1 - k_1 k_2 u_2 -k_1 k_3 u_3 & = 0,\nno\\
 u_2'' +\frac{b'}{b} u_2' + \frac{k_1^2}{b^2} u_1 +k_3^2 u_2 -\frac{k_1 k_2}{b^2} u_1 -k_2 k_3 u_3 &= 0,\label{eom_mode}\\
u_3'' +\frac{b'}{b} u_3' +\frac{k_1^2}{b^2} u_3 + k_2^2 u_3 -\frac{k_1k_3}{b^2} u_1 -k_2 k_3 u_2& = 0.\nno
 \end{align}
 Moreover, all the mode functions $u_1$, $u_2$, and $u_3$  satisfy the  constraint in Eq. \eqref{constraint_mode} which explicitly boils down to
 \beq\label{constraint_mode_mod}
 \frac{k_1}{b^2} u_1' + k_2 u_2' +k_3 u_3' =0.
 \eeq
The mode functions follow the normalization condition,

\beq\label{normalize_mod}
\left(u_i^{(p)}\tg^{jm}u_m^{*'(p)}-u_i^{*(p)}\tg^{jm}u_m^{'(p)}\right)=\frac{i}{2b}\left(\delta_i^j-\frac{k_ik^j}{k_nk^n}\right).
\eeq
Utilizing the above formalism of quantization along with the constraint relation, we evolve the mode function in different phases of the universe until the present epoch.

\section{Evolution of Electromagnetic field during inflationary era}\label{sec3}
According to our model in Eq.\eqref{aniso_model}, the anisotropy in the spacetime exists only towards the end of inflation. After the end of inflation, within a short period, the spacetime essentially becomes FLRW again, as seen from Fig. \ref{aniso_model}. However, it is important to mention that the large-scale production of the EM field is not affected due to this short presence of anisotropy after inflation. It is also evident that $b\rightarrow 1$ towards past infinity ensures that the Bunch-Davis vacuum condition is satisfied in the infinite past. Furthermore, we assume the background spacetime is de Sitter in nature, i.e., $a=-{1}/{(H\eta)}$, where $H$ is the Hubble parameter during inflation and remains constant throughout the entire inflation. Following these initial conditions, we numerically solve the mode function equations shown in Eq. \eqref{eom_mode}. We consider $ k_1=k_2=k_3=\tk\sim 1 ~\mbox{Mpc}^{-1}$ for simplification and the Hubble parameter $H=10^{-5} M_{\rm pl}$ remains constant throughout the inflationary era.  Here $M_{\rm pl}=\sqrt{{1}/{(8\pi G)}}$ is the reduced Planck mass. We redefine the conformal time $\eta$  as a dimensionless parameter $x=\tk\eta$. In terms of this new variable, Eq. \eqref{model} can be rewritten as,
\beq\label{anisotropy_model}
b(x) =1+ \alpha e^{- \left(\frac{x}{x_m}\right)^2},
\eeq
where the parameters $\alpha$ and $\tk\eta_m$ are chosen accordingly to avoid the backreaction from anisotropy, which essentially means that the anisotropy acts as a perturbation over the FLRW universe. A detailed discussion of the anisotropic backreaction is done in the later section. In terms of the redefined variables $x=\tk \eta,\tilde{u}_i=\sqrt{\tk}u_i$, and $\tk=k_1=k_2=k_3$, the mode function equations can be written as,
\bea
\frac{d^2\tu_1}{dx^2}-\frac{1}{b}\frac{db}{dx}\frac{d\tu_1}{dx}+2\tu_1-\tu_2-\tu_3&=&0,\nno\\
\frac{d^2\tu_2}{dx^2}+\frac{1}{b}\frac{db}{dx}\frac{d\tu_2}{dx}+\frac{\tu_2-\tu_1}{b^2}-\tu_3&=&0,\label{eom_mode_mod}\\
\frac{d^2\tu_3}{dx^2}+\frac{1}{b}\frac{db}{dx}\frac{d\tu_3}{dx}+\frac{\tu_3-\tu_1}{b^2}-\tu_2&=&0,\nno
\eea
along with the constraint equation,
\beq
\frac{1}{b^2}\frac{d\tu_1}{dx}+\frac{d\tu_2}{dx}+\frac{d\tu_3}{dx}=0.
\eeq
By solving Eq. \eqref{eom_mode_mod}, we can obtain the mode function solution for different choices of the parameters $\alpha$ and $x_m $ as shown in Fig. \ref{mode_soln}.
\begin{figure}[t]
\begin{center}
\includegraphics[scale=0.65]{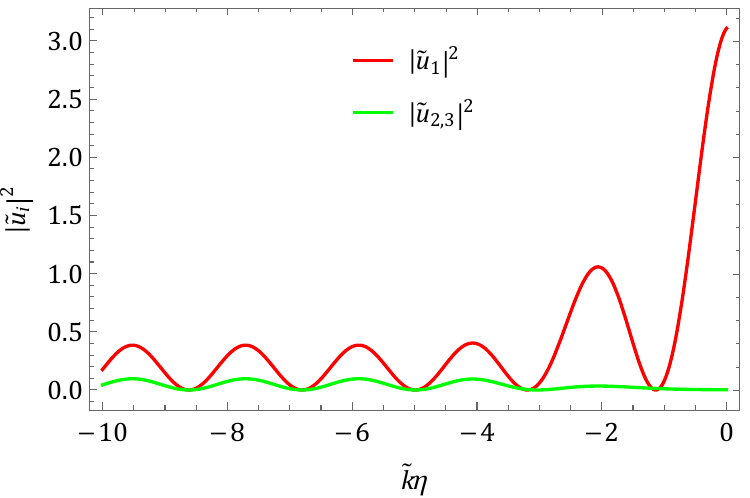}
\caption{Evolution of the mode functions $\tu_1$, $\tu_2$, and $\tu_3$ with $x=\tk\eta$ for the value of anisotropic parameter $\alpha=3$ and $ x_m = -2$. As the anisotropy exists only along the $x$-direction, the mode function equation corresponding to the $x$-direction $(\tu_1)$ behaves differently than the other two $(\tu_2,\tu_3)$.}\label{mode_soln}
\end{center}
\end{figure}
The above figure shows that the mode function grows in time due to anisotropy, particularly near the end of inflation. For values $\alpha < 0.03$, field production stops altogether. Hence, we get a lower bound on the anisotropic parameter {\bf$\alpha\geq0.03$}. The upper bound on $\alpha$ is discussed in the later sections.
\subsection{Power spectrum of the electromagnetic field during inflationary era}
 The stress energy-momentum tensor corresponding to the produced  EM field is given by
\beq
T_{\mu\nu}=-\frac{2}{\sqrt{-g}}\frac{\delta[\sqrt{-g}{\cal L}]}{\delta g^{\mu\nu}}.
\eeq
As a result, the energy-momentum tensor corresponding to the electromagnetic part of the Lagrangian boils down to
  \beq\label{energy_momentum_tensor}
  T_{mn}= -\frac{1}{4} g_{mn} g^{\mu\alpha} g^{\nu\beta} F_{\mu\nu}F_{\alpha\beta} + g^{\mu\nu} F_{m\mu} F_{n\nu}.
  \eeq
The total energy density of the system is given by the $T_{tt}$ component of the energy-momentum tensor. Therefore, the total electromagnetic energy density of the system is 
\beq\label{total_energy}
\rho=-\langle T^0_0\rangle = \frac{1}{2 a^2} g^{ij} \langle A_i' A_j'\rangle +\frac{1}{4} g^{ij} g^{ab} \langle F_{ia} F_{jb}\rangle.
\eeq  
Thus, we have the electric field and magnetic field energy densities as,
\bea\label{elec_mag_energy_dens}
\rho_E(x,\eta) &=& \frac{1}{2 a^2}g^{ij} \langle A'_iA'_j\rangle ,\nno\\
\rho_B(x,\eta)&=&\frac{1}{4} g^{ij} g^{mn}\langle F_{ij}F_{mn}\rangle,
\eea
respectively, where the expectation values are taken with respect to the initial Bunch-Davies (BD) vacuum.
In the momentum space, these energy densities can be written as
\bea\label{EM_density}
\rho_E (k,\eta) &=&\frac{1}{2 a^4}\sum_p \int \frac{d^3k}{(2\pi)^3}u_i^{(p)}\tg^{ij}u_j^{*'(p)}\nno,\\
\rho_B (k,\eta) &=& \frac{1}{4a^4}\sum_p\int\frac{d^3k}{(2\pi)^3} \tg^{ij}\tg^{mn} \nno\\
&&\times \bigg[\bigg(k_ik_n u_m^{(p)}u_j^{*(p)}
-k_i k_ju_m^{(p)} u_n^{*(p)}\bigg)\nno\\
&&+\left(k_m k_j u_i^{(p)}u_n^{*(p)}-k_m k_n u_i u_j^{*(p)}\right)\bigg].
\eea
In order to determine the strength of the magnetic field in the present era,  we first define the power spectrum of the electromagnetic field as
\bea
{\cal P}_{E/B}(k,\eta)&=&\frac{\pr \rho_{E/B}}{\pr \ln k},
\eea
as already stated earlier, each polarization mode follows the same equation of motion. Therefore, all the polarization modes have equal contributions. Summing over all the polarization modes and using the assumption amplitude of all the momentum $k_1=k_2=k_3=\tk$ to be the same, we calculate the power spectrum of the electric and magnetic field as
\begin{align}\label{power_spec_E}
{\cal P}_E(\eta,\tk)=&~ \frac{\tk^3}{2\pi^2 a^4}\left(\frac{|u'_1(\eta)|^2}{b^2}+|u'_2(\eta)|^2+|u'_3(\eta)|^2\right),\\
{\cal P}_B(\eta,\tk)=&~\frac{\tk^5}{2\pi^2 a^4}\bigg[\frac{1}{b^2}\bigg(2|u_1|^2+|u_2|^2+|u_3|^2\nno\\
&-2 \Re(u_1u_2^*)-2\Re(u_1u_3^*)\bigg)\nno\\
&+\bigg(|u_2|^2+|u_3|^2-2\Re(u_2u_3^*)\bigg)\bigg].\label{power_spec_B}
\end{align}
With these forms of the power spectrum, our goal would be to calculate its strength at present. However, before that, we will calculate the condition for which the produced electromagnetic field should not back react to the background during inflation. 
\subsection{Backreaction of anisotropic background and generated EM field}
In the previous section, we have briefly discussed the backreaction and strong coupling problem of inflationary magnetogenesis. In a general large-scale gauge field production scenario, a scalar field is coupled to the EM field to break the conformal invariance. Depending on the choice of the coupling function, it is possible to have a strong coupling problem, and different scenarios have been discussed in the literature \cite{Sharma:2017eps, Demozzi:2009fu, Ferreira:2013sqa, Tasinato:2014fia}. For the sake of completeness, we discuss it here briefly. In order to have a sustainable production of the electromagnetic field during inflation, the coupling function is often chosen to be an increasing function of time. However, it needs to revert to unity to restore the regular Maxwellian electromagnetism at the end of inflation. Hence, it needs to be very small at the start of the inflationary era, so the effective charge of electrons will be very large, and we cannot treat the gauge field as a free field during the inflationary era. In this proposal, there is no such direct coupling between the inflaton field and the EM field. 
Therefore, we do not need to worry about the strong coupling issue in this scenario. However, we must ensure that the anisotropy energy density or the generated EM field does not jeopardize the inflation. To this extent, we calculate the energy density produced by the anisotropic background and get a lower bound on the anisotropic parameter $\tk\eta_m$ and $\alpha$ introduced in Eq. \eqref{anisotropy_model}. The energy-momentum tensor of the background $T_{\mu\nu}$ is dictated by the Einstein equation in terms of the Einstein tensor $G_{\mu\nu}$ as 
\beq
G_{\mu\nu}=8\pi G T_{\mu\nu}.
\eeq

In the case of Bianchi type I background as introduced in Eq. \eqref{metric}, the $00$ component of the Einstein tensor can be calculated as
\bea
G_{00}&=& \frac{a'(3a'b+2ab')}{a^2 b}.
\eea
Thanks to this result,  we can calculate the energy density corresponding to the anisotropic background. It turns out as
 \bea\label{back_energy}
 \rho_{\rm total}&=&-T^0_0=-\frac{1}{8\pi G}G^0_0=\frac{1}{8\pi G}\left(3\frac{a'^2}{a^4}+2\frac{a'}{a^3}\frac{b'}{b}\right)\nno\\
 &=&3 H^2 M_{\rm pl}^2+2H M_{\rm pl}^2\frac{b'}{ab},
 \eea
 where $H\equiv {a'}/{a^2}$ is the Hubble parameter in conformal time during the inflation, $a$ is the scale factor in de Sitter spacetime $(a=-{1}/{(H\eta)})$. From the dynamics of the inflaton field during inflation, we already know that the total energy of the inflaton field is given by $\rho_{\rm inf}=3H^2 M_{\rm pl}^2$. Therefore, the total background energy density in Eq. \eqref{back_energy} consists of two parts. The first part we call inflationary energy density, and the second part is the energy density due to anisotropy in the background,
\beq
\rho_{\rm inf}=3H^2 M_{\rm pl}^2,~\rho_{\rm anis}=2 H M_{\rm pl}^2\frac{b'}{ab}.
\eeq 
In our proposition, we have mentioned earlier that the anisotropy should act as a perturbation. Therefore, we must ensure that the anisotropic energy density must be much lower than the inflaton energy density. Furthermore, the electromagnetic energy density has to be lower than the anisotropic and inflaton energy densities. From the PLANCK data \cite{Akrami:2018odb}, we know that the temperature anisotropy in CMB  is $\frac{\Delta T}{T}\sim 10^{-5}$. If the anisotropic energy is closer to the perturbative limit towards the end of the inflationary era, it will not affect the CMB map, as observed by the PLANCK.
We define e-folding number during the inflationary era as
$N=\ln\left(\frac{a}{a_{\rm end}}\right)$, where $a_{\rm end}$ is the scale factor at the end of inflation. By this definition, the e-folding number at the end of inflation $N_{\rm end}=0$. Moreover, the total e-folding number during the inflation is $N_{\rm tot} \simeq 60$.
 The anisotropic factor $b$ in terms of e-folding number can be written as
 \beq\label{aniso_model_efold}
 b(N)= 1 +\alpha \exp\left[-e^{2(N_m-N)}\right],
 \eeq
 here $N_m$ is the e-folding number corresponding to the conformal time $\eta_m$.
 We can calculate the ratio of the anisotropic energy density and the inflationary energy density in terms of the e-folding number $N$ as follows
\bea
\bigg\vert\frac{\rho_{\rm anis}}{\rho_{\rm inf}}\bigg\vert
=\bigg|\frac{2}{3 H}\frac{b'}{ab}\bigg|
= \bigg|\frac{2}{3}\frac{db}{dN}\frac{1}{b}\bigg|.
\eea
\begin{figure}[t]
    \centering
    \includegraphics[scale=0.65]{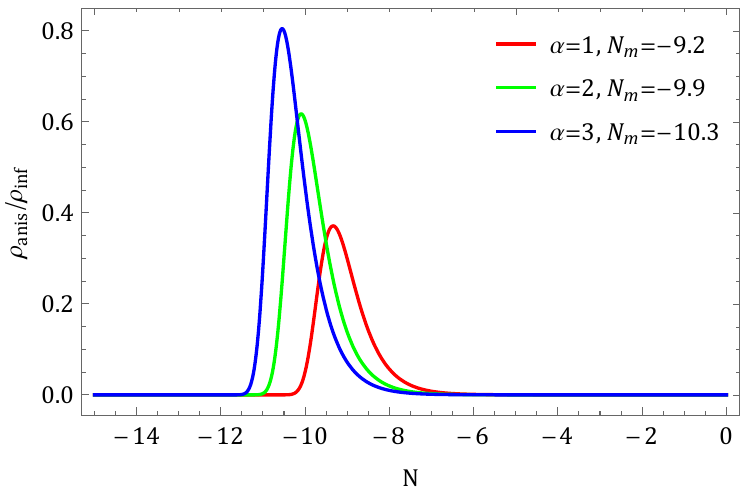}
    \caption{Evolution of the ratio of anisotropic energy density to inflationary energy density with the e-folding number $N$ with different anisotropic parameters $\alpha$ and  $N_m$. }
    \label{anis_vs_inf}
\end{figure}  
In order to have sustainable inflation, such that the anisotropic energy density does not affect the inflation energy density, we need to have  $\bigg|
\frac{\rho_{\rm anis}}{\rho_{\rm inf}}\bigg|< 1$ throughout the entirety of the inflation. Thus, the ratio gives us an upper bound on $\alpha$, which dictates the strength of the anisotropy. In Fig. \ref{anis_vs_inf}, we can see that the ratio of the energy densities reaches its maximum towards the end of inflation. Thus, we can choose our parameters such that the ratio is up to the perturbative level $(\sim 0.5)$. It gives us the upper bound $\alpha\leq 1.48$. Still, the CMB remains unaffected due to the presence of spatial anisotropy. However, it is worth mentioning here that we do not consider the dynamics of the anisotropy in the background. In order to make sure that the anisotropic background comes in towards the end of inflation, we take the upper limit on the parameter $\tk\eta_m\geq -2 $. 
 Furthermore, during inflation, the EM field also gets produced. It is also necessary to ensure that the generated gauge field energy density does not violate the inflationary energy density. 
We can see that the maximum production occurs towards the end of inflation from the nature of the coupling function introduced in Eq. \eqref{anisotropy_model}. Thus, to avoid the backreaction problem, it is sufficient to satisfy 
\beq
 \rho_E+\rho_B\leqslant\rho_{\rm inf}.\nno
\eeq
We can obtain the values of the energy densities of the electric and magnetic fields from Eq. \eqref{elec_mag_energy_dens} and integrate over all the modes inside the horizon during the inflationary era. Which finally boils down to,
\bea
\rho_E
&=&\frac{H^4}{2\pi^2}\int_{x_i}^{x_f} dx x^3 \bigg[\frac{1}{b^2}\bigg|\frac{d\tu_1}{dx}\bigg|^2\nno\\
&&+\bigg|\frac{d\tu_2}{dx}\bigg|^2+\bigg|\frac{d\tu_3}{dx}\bigg|^2\bigg],\\
\rho_B
&=&\frac{H^4}{\pi^2}\int_{x_i}^{x_f}dx x^3\bigg[\frac{1}{b^2}\bigg(2|\tu_1|^2+|\tu_2|^2+|\tu_3|^2\nno\\
&&-2\Re(\tu_1 \tu_2^*)-2\Re(\tu_1 \tu_3^*)\bigg)\nno\\
&&+\bigg(|\tu_2|^2+|\tu_3|^2-2\Re(\tu_2 \tu_3^*)\bigg) \bigg].
\eea 
Where we recall the variables $x=\tk\eta,\tu_i=\sqrt{\tk}u_i$, evaluating the integrations numerically, the ratio of the energy densities turns out to be $\frac{\rho_{\rm E}+\rho_{\rm B}}{\rho_{\rm inf}}\sim 10^{-9}$, for the anisotropic parameter $\tk\eta_m=-1$ and $\alpha=1.45$. As the generated electromagnetic energy density is very low compared to the background inflaton energy density, the backreaction problem is also avoided. Therefore, with this formalism, we can sustainably produce the EM field during inflation without worrying about the strong coupling or backreaction problem. On the other hand, ensuring that the generated EM field does not surpass the energy density of the inflationary background is also necessary. Eq. \eqref{anisotropy_model} shows that the maximum energy density occurs at $\eta=0 $. However, we have taken that the inflation ends at $\eta_f$, so there is no production of the large-scale magnetic field in the post-inflationary era. If the electromagnetic energy density is less than the anisotropic energy density at the end of inflation, all the sufficient conditions for no backreaction are satisfied. To this end, we reiterate that the anisotropic parameter $\alpha$ is so chosen that the anisotropic energy density remains subdominant compared to the inflaton energy density. We further show that the produced energy density of the electromagnetic field is less than the anisotropic energy density. 
In conclusion, the produced electromagnetic field affects neither the inflationary nor the anisotropic background. Therefore, this formalism effectively produces a magnetic field without special coupling to avoid the backreaction effect.
\section{Post inflationary evolution}\label{sec4}
The anisotropic factor $b$ goes to unity after the end of inflation, and the spacetime becomes conformally flat. The EM field evolves as a usual Maxwellian field subsequently. However, depending on the evolution of the universe, we can have two different scenarios of field evolution: (i) In the first scenario, it is assumed that the universe instantly goes into radiation domination, i.e., the inflaton field instantly decays and produces radiation. (ii) In the second scenario, the inflaton decays within a finite time, and therefore, it goes through a brief period of reheating era having a non-zero e-folding number and very low conductivity. The dynamics of the subsequent evolution of the universe dictate the present strength of the observed magnetic field. We will discuss both scenarios in the next subsections.
\subsection{The case of instantaneous reheating}
Here in this section, we will find the strength of the magnetic field in the present time, considering an instantaneous reheating scenario. In this case, after the end of the inflationary era, the universe instantly thermalizes and goes to the radiation-dominated era. As the conductivity of the universe becomes very large, the electric field dies out instantly. However, the magnetic field produced during the inflationary era decays as a radiation density ${\cal P}_B\propto a^{-4}$. Therefore, incorporating the conservation of entropy, we can compute the strength of the magnetic field, relating to the field strength at the end of inflation. We have already calculated the power spectra of the magnetic field during the inflationary era in Eq. \eqref{power_spec_B}. In terms of the mode functions, the explicit expression for the present-day magnetic field turns out as
\beq\label{present_magnetic_str}
B_0
=\left(\frac{-\tk\eta_f a_f  H}{\pi^2 a_0}\right)^2 \sqrt{|\tu_1(x_f)|^2+|\tu_2(x_f)|^2+|\tu_3(x_f)|^2}.
\eeq
The above expression is in the $\mbox{GeV}^2$ unit. Here, we recall the variable $\tu_i=\sqrt{\tk}u_i$ (with $i=1$, $2$, $3$), $H$ is the Hubble parameter during inflation, and $\tk$ is the scale under consideration in which we will estimate the strength of the magnetic field. Furthermore, in order to calculate the value of $B_0$ from Eq. \eqref{present_magnetic_str}, we first need to evaluate the value of the ratio $ \frac {a_f}{a_0}$. We evaluate the value to be $\frac{a_0}{a_f}\approx 10^{30}(H/10^{-5}M_{\rm pl})^{1/2}$. Here, in particular, we have taken the value of the Hubble parameter to be $H=10^{-5} M_{\rm pl}$. With the numerical solution of the mode functions from Fig. \ref{mode_soln} at the end of inflation $\tk \eta_f=-0.0001$ and Eq. \eqref{present_magnetic_str} we can evaluate the strength of the magnetic field at present-day using the conversion $1~\mbox{G}= 1.95\times 10^{-20} ~\mbox{GeV}^2$ for different values of the anisotropic parameters $\alpha$ and $\tk\eta_m$.
\begin{figure*}[t]
    \begin{center}
     \subfigure[]{\includegraphics[scale=0.53]{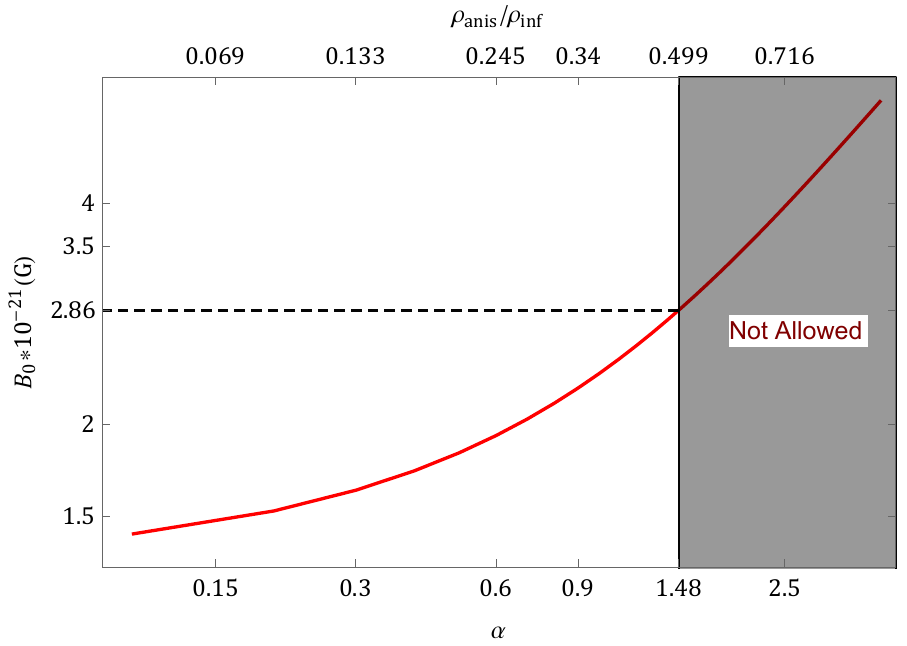}}
  \subfigure[]{\includegraphics[scale=0.63]{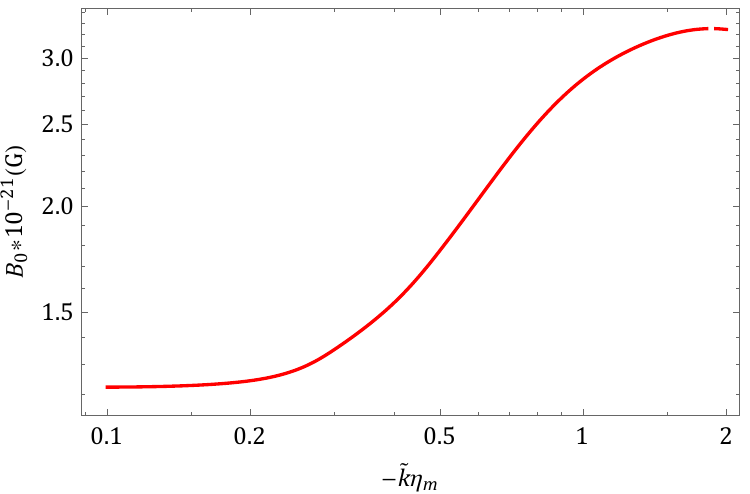}}
    \caption{\textbf{(a)} Variation of the present strength of the magnetic field $B_0$ with $\alpha$ for a fixed value of the anisotropic parameter $\tk\eta_m=-1$. \textbf{(b)} Variation of the present strength of the magnetic field $B_0$ with $\tk\eta_m$ for a fixed value of the anisotropic parameter $\alpha=1.45$, the ratio of the anisotropic energy density in this case remains constant at $\rho_{\rm anis}/\rho_{\rm inf}=0.492$.  }
    \label{B0_vs_anm}
    \end{center}
\end{figure*}

In Fig. \ref{B0_vs_anm}, we can see the variation of the present-day magnetic field $B_0$ with $\alpha$ for a fixed value of $\tk\eta_m$ as well as the variation of magnetic field strength with $\tk\eta_m$. The maximum value of $B_0$ obtained in the instant reheating scenario for fixed value of $\tk\eta_m$ is $B_0=2.86\times 10^{-21}~\mbox{G}$ varying $\alpha$. Similarly for a fixed value of $\alpha$, the maximum value of $B_0$ obtained is $B_0 = 3.24\times 10^{-21}~G$. 
Experiments like Faraday rotation and gamma-ray observation impose a bound on the present strength of primordial magnetic field  $10^{-10}~\mbox{G}\lesssim B_0 \lesssim 10^{-22}~\mbox{G}$ \cite{Durrer:2013pga}. Therefore, this proposal can generate a large-scale magnetic field within the experimental bound of the present-day intergalactic magnetic field. With the choice of the anisotropic parameter in the range $0.03\leq \alpha\leq 1.48$, the backreaction or the strong coupling problem is also avoided.
\subsection{The case of prolonged reheating with constant equation of state} 
In the last section, we saw that we can generate the required strength of the magnetic field in the instant reheating scenario. However, if we consider a reheating phase with a non-zero e-folding number, then the conductivity of the universe does not reach infinity instantly. Instead, during this period, the conductivity of the universe may remain to be very low. As a result, the electric field does not go to zero immediately and induces a magnetic field during this period. This conversion of the electric field into the magnetic field during the reheating phase occurs through Faraday induction \cite{Kobayashi:2019uqs}. This conversion of an electric field to a magnetic field makes it diluted slowly during the reheating era compared to the previous case of ${\cal P}_B\propto a^{-4}$. Thus, a finite reheating era further strengthens the magnetic field on a large scale and gives us bounds on the EoS during the reheating era. After the inflation ends, the anisotropic factor $b$ goes to unity after a very short period, and the EM field evolves in the usual manner. Following the regular Maxwellian evolution, the equation of motion of the mode functions $u_i(\tk,\eta)$ during the reheating becomes
\beq\label{mode_reh}
u_i^{''(re)} (\tk,\eta)+ 3\tk^2  u_i^{(re)}(\tk,\eta)=0,
\eeq
where $i=1,~2,~3$ are the indices corresponding to three spatial components of the gauge field and $u_i^{(re)}(\tk,\eta)$ are the mode functions during reheating. Furthermore, we consider the universe a poor conductor during this period. To be precise, we take the conductivity to be zero. The solution of the mode function from Eq. \eqref{mode_reh}, along with the proper normalization condition, gives us
\bea\label{mode_soln_reh}
u_1^{(re)}(\tk,\eta) &=& \frac{1}{\sqrt{6\sqrt{3}\tk}}\bigg[\alpha_1(\tk) e^{-i\sqrt{3} \tk(\eta-\eta_f)}\nno\\
&&+\beta_1(\tk) e^{i\sqrt{3} \tk(\eta-\eta_f)}\bigg],\nno\\
u_{2,3}^{(re)}(\tk,\eta) &=& -\frac{1}{2\sqrt{6\sqrt{3}\tk}}\bigg[\alpha_{2,3}(\tk) e^{-i\sqrt{3}\tk (\eta-\eta_f)}\nno\\
&&+\beta_{2,3} (\tk)e^{i\sqrt{3} \tk(\eta-\eta_f)}\bigg],
\eea
with $\alpha_i$ and $\beta_i$ are the integration constants, and $\eta_f$ denotes the end of inflation. 
the integration constants are evaluated at the end of inflation $\eta_f$ by equating the junction conditions of inflationary and reheating era
\beq\label{junction_cond}
u_i^{(re)}(\tk,\eta_f)=u_i(\tk,\eta_f) ~~\text{and}~~u_i^{'(re)}(\tk,\eta_f)=u_i'(\tk,\eta_f).
\eeq
In the above Eq. \eqref{junction_cond}, $u_i(\tk,\eta)$ are mode functions during the inflationary era which follow Eq. \eqref{eom_mode} and $u_i^{(re)}$ are mode functions during the reheating era following Eq. \eqref{mode_reh}.  This immediately leads to the integration constants, 
\begin{align}
\alpha_1(\tk)&=\sqrt{\frac{3\sqrt{3}\tk}{2}}u_1(\tk,\eta_f) + i\sqrt{\frac{\sqrt{3}}{2\tk}}u_1'(\tk,\eta_f),\nno\\
\beta_1 (\tk) &= \sqrt{\frac{3\sqrt{3}\tk}{2}}\tu_1(\tk,x_f) - i\sqrt{\frac{\sqrt{3}}{2\tk}}u_1'(\tk,\eta_f),\nno\\
\alpha_{2,3}(\tk) &= -\sqrt{6\sqrt{3}\tk}u_{2,3}(\tk,\eta_f)-i\sqrt{\frac{2\sqrt{3}}{\tk}}u'_{2,3}(\tk,\eta_f),\nno\\
\beta_{2,3} (\tk)&=-\sqrt{6\sqrt{3}\tk}u_{2,3}(\tk,\eta_f)+i \sqrt{\frac{2\sqrt{3}}{\tk}}u'_{2,3}(\tk,\eta_f).
\end{align}
With all these, we can now compute the time-evolving power spectrum during reheating as
\bea\label{pow_spec_reh}
{\cal P}_B(\eta,\tk)&=& \frac{\tk^5}{\pi^2 a^4}\left(|u_1^{(re)}|^2+|u_2^{(re)}|^2+|u_3^{(re)}|^2\right)\nno\\
&=& \sum_i \frac{\tk^4}{\pi^2 a^4}|\tu_i^{(re)}|^2\nno\\
&=& \frac{\tk^4}{\pi^2 a^4}\bigg[ \frac{1}{6\sqrt{3}}\bigg(|\alpha_1|^2+|\beta_1|^2+ 2|\alpha_1||\beta_1|\nno\\
&&\times \cos[Arg(\alpha_1\beta_1^*) -2\sqrt{3}\tk(\eta-\eta_f)]\bigg)\nno\\
&+& \frac{1}{24\sqrt{3}}\bigg(|\alpha_2|^2+|\beta_2|^2+ 2|\alpha_2||\beta_2| \nno\\
&&\times \cos[Arg(\alpha_2\beta_2^*) -2\sqrt{3}\tk(\eta-\eta_f)]\bigg)\nno\\
&+& \frac{1}{24\sqrt{3}}\bigg(|\alpha_3|^2+|\beta_3|^2+ 2|\alpha_3||\beta_3|\nno\\
&& \times \cos[Arg(\alpha_3\beta_3^*) -2\sqrt{3}\tk(\eta-\eta_f)\bigg)\bigg].
\eea
In order to estimate the strength of the magnetic field during the reheating era, we first need to evaluate the term $\eta-\eta_f$. Following Ref. \cite{Kobayashi:2019uqs} the term is calculated as
\beq\label{reheat_term}
\eta-\eta_f =\int_{a_f}^a \frac{da}{a^2 H}.
\eeq
As the Hubble constant $H$ is present in the above equation, it is evident that the quantity $\eta-\eta_f$ depends on the background's evolution during the inflationary era. In particular, how the inflaton energy density is converted into radiation energy density. In general, there are two scenarios 
\begin{itemize}
    \item Evolution through time-independent, effective equation of state.
    \item 
    Perturbative decay of inflaton into radiation (perturbative reheating scenario).
\end{itemize}
Here in this paper, we will only discuss evolution through an independent constant effective EoS. In this context, we follow the methodology proposed by Kamionkowski et al. in Ref. \cite{Dai:2014jja}. Here, the evolution of the background is parametrized by a constant effective EoS $\omega_{eff}$. Therefore, the Hubble parameter during the reheating evolves as $H\propto a^{-\frac{3}{2}(1+\omega_{eff})}$. The physical parameters of reheating, like the e-folding number of the reheating era $N_{re}$ and the reheating temperature $T_{re}$, can be expressed in terms of the inflationary parameters and effective EoS $\omega_{eff}$ as \cite{Cook:2015vqa}
\bea\label{Nre}
N_{re}&=&\frac{1}{3\omega_{\rm eff}-1}\bigg[\ln(\rho_{f})-\ln\left(\frac{\pi^2 g_{re}}{30}\right)-\frac{1}{3}\ln\left(\frac{43}{11 g_{s,re}}\right)\nno\\
&&-4\ln\left(\frac{a_0T_0}{k}\right)+4\ln(H_k) +4N_k \bigg],
\eea
\beq\label{Tre}
T_{re}= \left(\frac{43}{11 g_{s,re}}\right)^{1/3}\left(\frac{a_0T_0}{k}H_k e^{-N_k}e^{-N_{re}}\right),
\eeq
where $H_k$ denotes the Hubble parameter at the time of horizon crossing, $k/a_0=0.05 ~\mbox{Mpc}^{-1}$ is the pivot scale, $g_{re}$ is the degrees of freedom during reheating and $N_k$ is the total e-folding number from the end of inflation till horizon crossing. As we have not considered any particular inflation potential in this paper, we develop a model-independent way to determine $N_k$ following Ref.\cite{Maity:2021qps}. In
the calculation of $N_k$ (see Appendix \ref{appen3}), we have taken
the central values of 
 scalar spectral index $n_s=0.9649$ and scalar perturbation amplitude $\ln[10^{10}{\cal A}_s]=3.044$, considering the constraints provided by the PLANCK data \cite{Akrami:2018odb} and as an input parameter we have chosen $N_k=50$. With this choice of $n_s, N_k$, we get an upper bound on the effective EoS $\omega_{eff}<0.164$ from the BBN bound of reheating temperature $T_{re}\sim 10^{-2}$ GeV.
Now, in order to connect the reheating parameters $N_{re}, T_{re}$ to the strength of the primordial magnetic field, we need to evaluate the quantity $\eta-\eta_f$ in Eq. \eqref{reheat_term}. It is evaluated following the evolution of the Hubble parameter during the reheating era. As the EoS is constant $\omega_{eff}$, the variation of the Hubble parameter during the reheating era $(H_{re})$ is related to the Hubble parameter at the end of inflation $(H_f)$ as
\beq
H_{re}= H_f \left(\frac{a_{re}}{a_f}\right)^{-\frac{3}{2}(1+\omega_{eff})},
\eeq
where the subscript ``$re$'' represents the end of reheating. Thus, $a_{re}$ and $H_{re}$ are the scale factor and Hubble parameter at the end of reheating, respectively.
Following the above relation, the term in Eq. \eqref{reheat_term} boils down to
\beq
\eta -\eta_f = \frac{2}{1+ 3 \omega_{eff}}\left(\frac{1}{a H}-\frac{1}{a_f H_f}\right).
\eeq
Substituting the value of the extra reheating term $\eta-\eta_f$, we can calculate the present strength of the magnetic field as a function of the effective EoS $\omega_{eff}$.
 After the end of reheating, the conductivity of the universe goes to infinity. Therefore, the electric field goes to zero, and the Faraday conversion of the electric field into the magnetic field stops at the end of reheating. And the magnetic field decays as radiation $(a^{-4})$ until now. From the conservation of magnetic energy density, the present strength of the magnetic field can be calculated from the relation as follows
\beq\label{mag_present_reh}
\frac{\pr \rho_B}{\pr \ln k}\bigg\vert_0=\left(\frac{a_{re}}{a_0}\right)^4 \frac{\pr \rho_B}{\pr\ln k}\bigg\vert_{\rm re}.
\eeq 
Evolving through the reheating era, the strength of the magnetic field in the present era turns out as
\bea\label{mag_pres_reh}
B_0 &=&\frac{\sqrt{2}}{6\pi\sqrt{3}}\left(\frac{\tk}{a_0}\right)^2\Bigg[{\cal I}_1 + \frac14 \left({\cal I}_2 + {\cal I}_3 \right) \Bigg]^{1/2},
\eea
where,
\bea
{\cal I}_i &=& |\alpha_i|^2+|\beta_i|^2+ 2|\alpha_i|~|\beta_i|\cos(\mbox{Arg}(\alpha_i\beta_i^*)- \Phi),\nno\\
\Phi &=& \frac{4\tk\sqrt{3}}{(1+3 \omega_{eff})a_f H_f}\left(\left(\frac{H_f}{H_{re}}\right)^{\delta}-1\right).
\eea
In the above Eq. \eqref{mag_pres_reh}, $\delta= (3\omega_{eff}+1)/ (3\omega_{eff}+3)$. Varying the EoS $\omega_{eff}$, we get the present-day strength of the magnetic field of $B_0 \sim 4\times 10^{-20}$ G, which is one order higher than what was predicted for instantaneous reheating case, which is $\sim 3\times 10^{-21}$ G. Furthermore, from the observed strength of the magnetic field, we also get a lower bound of EoS $\omega_{eff}>0.132$. 
\begin{figure}[t]
\begin{center}
    \includegraphics[scale=0.6]{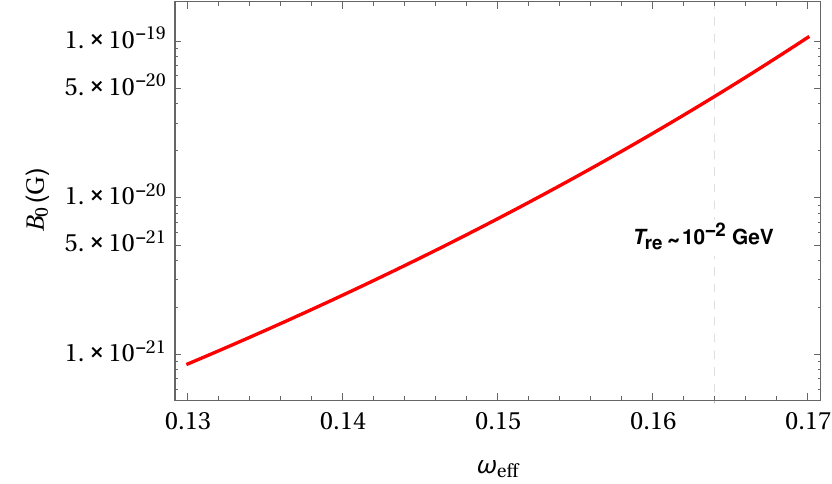}
    \caption{Variation of present magnetic strength with effective equation of state $\omega_{eff}$ for the choice of anisotropic parameters $\alpha=1.45$ and $\tk\eta_m=-1$. With the model independent formalism input parameter $N_k=50$ and inflationary parameter $n_s=0.9649$.} \label{b_vs_w}
\end{center}   
\end{figure}
 From Fig. \ref{b_vs_w}, we see that the present strength of the magnetic field increases due to the Faraday conversion of the electric field into the magnetic field; such increment is quite insensitive to the reheating EoS. 
 This increment is small since the strength of the electric field compared to the magnetic field at the end of inflation is not significantly higher. 
\section{Summary and Conclusions}\label{sec5}
This paper proposes a new formalism to generate large-scale magnetic fields during the inflationary era. The novelty of the work lies in the generation of fields during inflation. Several works have been done in the context of inflationary magnetogenesis. However, all the previous works rely on the conformal breaking coupling of the EM field with some scalar field or gravity. In the present case, we have taken the underlying background to be an anisotropic one (Bianchi type I), keeping conformal property intact.
Due to this, our model does not suffer from the usual strong coupling problem. In the process, we have introduced two parameters $\alpha$ and $\eta_m$ to characterize the behavior of the anisotropic scale factor $b(\eta)$. By appropriately tuning those anisotropic parameters, we further addressed the backreaction problem. If we take the ratio of the anisotropic energy density $\rho_{\rm anis}$ and the inflaton energy density $\rho_{\rm inf}$ to be $\rho_{\rm anis}/\rho_{\rm inf}\leqslant 0.5$ we get an upper bound of $\alpha\leq1.48$.
Furthermore, to ensure that electromagnetic field gets produced during the inflationary era, we get a lower bound on the parameter $\alpha\geq 0.03$. The parameter $\eta_m$ is so chosen that the anisotropy appears towards the end of inflation. For this, we have taken $\tk\eta_m\geq -2$, ensuring that the anisotropy is localized and short-lived. With this choice of parameters, we find that the ratio of the energy density of the generated EM field to the total inflaton energy density is $\sim 10^{-9}$, which implies that the electromagnetic energy density is also lower than the anisotropic energy density. Therefore, the generated electromagnetic field neither back-reacts on the inflaton field nor the anisotropic background. Finally, this set of parameters gives us a present strength of magnetic field $B_0\sim 3\times 10^{-21}~G$, for $\alpha=1.45$ and $\tk\eta_m=-2$, which is well in between the latest bound on present-day magnetic field strength. However, if we consider an elongated reheating period followed by inflation, the magnetic field strength further increases. This increase in strength occurs due to Faraday's conversion of the electric field to the magnetic field. By this prolonged reheating era, we get the present strength of magnetic field $B_0\sim 4\times 10^{-20} ~G$.  Through the introduction of the reheating era, we also get a tight constraint on the range of equation of state $0.132<\omega_{eff}<0.164$ for the particular choice of inflationary parameter $n_s$ and $N_k$. Due to the presence of anisotropy, there might be interesting signatures of the anisotropy on gravitational waves at small scales. Further, the most interesting would be investigating the origin of such anisotropy, particularly near the end of inflation. All those questions we leave for our future study. 
\noindent
\begin{acknowledgments}
DM wishes
to acknowledge support from the Science and Engineering Research Board (SERB), Department of Science, and
Technology (DST), Government of India (GoI), through the Core Research Grant CRG/2020/003664. DM and SP also thanks the Gravity and High Energy
Physics groups at IIT Guwahati for illuminating discussions.
\end{acknowledgments}

 \begin{appendix}
\begin{widetext}

\section{Power spectrum of the electromagnetic field during inflationary era}
 We have the energy-momentum tensor corresponding to the free Maxwellian Lagrangian, 
  \bea
  T_{mn}
  = -\frac{1}{4} g_{mn} g^{\mu\alpha} g^{\nu\beta} F_{\mu\nu}F_{\alpha\beta} + g^{\mu\nu} F_{m\mu} F_{n\nu}.
  \eea
  The energy density of the electromagnetic field is obtained from the "00" component of the energy-momentum tensor, which boils down to
\bea
T_{00} = \frac{1}{2} g^{ij} A_i' A_j' +\frac{a^2}{4} g^{im} g^{jn} F_{ij} F_{mn} .
\eea
Upon trading the EM field into the quantum operator and referring to Eq.\eqref{elec_mag_energy_dens}, we have the expression for the electric and magnetic field energy densities as
\bea\label{em_dens}
\rho_E(x,\eta) &=& \frac{1}{2 a^2}g^{ij} \langle A'_iA'_j\rangle ,\nno\\
\rho_B(x,\eta)&=&\frac{1}{4} g^{ij} g^{mn}\langle F_{ij}F_{mn}\rangle,
\eea
where the expectation value is obtained over the BD vacuum.
The expectation value of $g^{im}g^{jn}F_{ij}F_{mn}$ in the BD vacuum in terms of mode functions boils down to
\bea
\langle g^{im}g^{jn}F_{ij}F_{mn}\rangle &=&\sum_p \int \frac{d^3\tk}{(2\pi)^3}
\bigg[\frac{2\tk^2}{a^4b^2}\left(2|u_1|^2 +|u_2|^2+|u_3|^2-u_1 u_2^* -u_2 u_1^*-u_1 u_3^*-u_3u_1^*\right)\nno\\
&+&\frac{2\tk^2}{a^4}\left(|u_2|^2+|u_3|^2 -u_2 u_3^*-u_3 u_2^*\right)\bigg]\nno\\
&=& \sum_p \int \frac{d^3\tk}{(2\pi)^3}
\bigg[\frac{2\tk^2}{a^4b^2}\left(2|u_1|^2 +|u_2|^2+|u_3|^2-2 \Re(u_1 u_2^*)-2 \Re(u_1 u_3^*)\right)\nno\\
&+& \frac{2\tk^2}{a^4}\left(|u_2|^2+|u_3|^2 -2 \Re (u_2 u_3^*)\right)\bigg].
\eea
Substituting the expectation value of the term $g^{im}g^{jn}F_{ij}F_{mn}$ in Eq.\eqref{em_dens}, we get the energy density of the magnetic field. With the polarization index, the expression for the magnetic field energy density turns out as
\bea
\rho_B (\tk,\eta) &=&  \sum_p\int \frac{d^3\tk}{(2\pi)^3}\frac{\tilde{k}^2}{2a^4}\bigg[\frac{1}{b^2}\left(2 |u_1^{(p)}|^2+|u_2^{(p)}|^2+|u_3^{(p)}|^2-2 \Re(u_1^{(p)} u_2^{*(p)})-2\Re(u_1^{(p)} u_3^{*(p)})\right)\nno\\
&+&\left(|u_2^{(p)}|^2+|u_3^{(p)}|^2-2\Re(u_2^{(p)} u_3^{*(p)})\right)\bigg].
\eea
As all the polarization modes behave the same way, summing over all the polarization, we finally get the energy density of the magnetic field
\bea
\rho_B (\tk,\eta) &=&  \int \frac{d^3\tk}{(2\pi)^3}\frac{\tilde{k}^2}{a^4}\bigg[\frac{1}{b^2}\left(2 |u_1|^2+|u_2|^2+|u_3|^2-2 \Re(u_1 u_2^*)-2\Re(u_1 u_3^*)\right)\nno\\
&+&\left(|u_2|^2+|u_3|^2-2\Re(u_2 u_3^*)\right)\bigg].
\eea
Similarly, evaluating the expectation value of the term $g^{ij}A_i'A_j'$ and substituting it back in Eq.\eqref{em_dens}, we get the energy density of the electric field in terms of mode functions as
\beq
\rho_E (\tk,\eta) =  \int \frac{d^3\tk}{(2\pi)^3}\frac{1}{a^4}\left(\frac{|u_1'|^2}{b^2}+|u_2'|^2+|u_3'|^2\right).
\eeq
\section{Backreaction of anisotropic background and generated electromagnetic field}
 The energy-momentum tensor of the background $T_{\mu\nu}$ is dictated by the Einstein equation
\beq
G_{\mu\nu}=8\pi G T_{\mu\nu},
\eeq
where $G_{\mu\nu}$ is the Einstein tensor and $G$ is the gravitational constant. The Einstein tensor can be calculated in terms of the Riemann tensor $(R_{\mu\nu})$ and Ricci scalar $(R)$,
\bea
G_{\mu\nu}&=&R_{\mu\nu}-\frac{1}{2}R g_{\mu\nu}.\nno
\eea
In the case of Bianchi type I background as introduced in Eq. \eqref{metric}, the ``00" component of the Einstein tensor turns out as
\bea
G_{00}&=& \frac{a'(3a'b+2ab')}{a^2 b}.
\eea
This essentially gives us the background energy density,
 \bea
 \rho_{total}=-T^0_0 =\frac{1}{8\pi G}\left(3\frac{a'^2}{a^4}+2\frac{a'}{a^3}\frac{b'}{b}\right)=3 H^2 M_{pl}^2+2H M_{pl}^2\frac{b'}{ab}.
 \eea
 The ratio of anisotropic energy density to the inflaton energy density is given in terms of the e-folding number $N$,
\bea
\bigg|\frac{\rho_{anis}}{\rho_{inf}}\bigg| &=&\frac{2 H M_{pl}^2\frac{b'}{a b}}{3 H^2 M_{pl}^2 }
=\frac{2}{3}\frac{db}{dN}\frac{1}{ b},\nno
\eea
with the e-folding number is defined as $dN= d\ln a$, where $a$ is the scale factor. From the above equation, we can get the ratio of the anisotropic and the inflationary energy densities.\\
The backreaction problem can be evaded if the total energy of the generated EM field is less than the energy density of the inflaton field, that is  \beq
 \rho_E +\rho_B \leqslant \rho_{inf} .
 \eeq
 The total energy densities of the EM are given by Eq.\eqref{EM_density}. Integrating all the modes, we get the total energy density. The modes involved are given by $\tk_i=a_i H$, which crosses the horizon at the beginning of inflation, and $k_f=a_f H$ are the modes that cross the horizon at the end of inflation. We have the total energy density of the electric field expressed as,
 \bea\label{elec_density}
 \rho_E&=&\int_{\tk_i}^{\tk_f} \frac{d\tk}{\tk}\frac{\tk^3}{2\pi^2a^4}\bigg(\frac{|u_1'(\eta)|^2}{b(\eta)^2}+|u_2'(\eta)|^2+|u_3'(\eta)|^2\bigg)\nno\\
 &=&\frac{1}{2\pi^2}\int_{\tk_i}^{\tk_f}d\tk\frac{\tk^3}{a^4}\bigg(\frac{1}{b^2}\bigg|\frac{d\tu_1}{dx}\bigg|^2+\bigg|\frac{d\tu_2}{dx}\bigg|^2+\bigg|\frac{d\tu_3}{dx}\bigg|^2\bigg)\nno\\
 &=&\frac{H^4}{2\pi^2}\int_{\tk_i}^{\tk_f}d(\tk\eta)(\tk^3\eta^3)\bigg(\frac{1}{b^2}\bigg|\frac{d\tu_1}{dx}\bigg|^2+\bigg|\frac{d\tu_2}{dx}\bigg|^2+\bigg|\frac{d\tu_3}{dx}\bigg|^2\bigg)\nno~~\mbox{(Substituted $a=-1/H \eta$)}\\
 &=&\frac{H^4}{2\pi^2}\int_{x_i}^{x_f}dx x^3\bigg(\frac{1}{b^2}\bigg|\frac{d\tu_1}{dx}\bigg|^2+\bigg|\frac{d\tu_2}{dx}\bigg|^2+\bigg|\frac{d\tu_3}{dx}\bigg|^2\bigg) .
 \eea
 Similarly, the total energy density of the magnetic field is calculated
 \bea
 \rho_B=\frac{H^4}{2\pi^2}\int_{x_i}^{x_f}dx x^3\bigg[\frac{1}{b^2}\bigg(2|\tu_1|^2+|\tu_2|^2+|\tu_3|^2-2 \Re(\tu_1\tu_2^*)-2\Re(\tu_1\tu_3^*)\bigg)
+\bigg(|\tu_2|^2+|\tu_3|^2-2\Re(\tu_2\tu_3^*)\bigg)\bigg],
 \eea
 where $x=\tk\eta$ and $\tu_i=\sqrt{\tk}u_i$. Integrating over the limits numerically with the solutions of the mode functions, we get the total energy density of the generated EM field. Finally, comparing the energy density of the EM field to the inflaton field $(\rho_{inf}=3 H^2 M_{pl}^2)$, we get
 \beq
 \frac{\rho_E+\rho_B}{\rho_{inf}}\sim 10^{-9}.
 \eeq
 After the inflation, the production of electromagnetic field stops altogether. If we consider an instant reheating scenario, it essentially behaves as a radiation field. Thus, by conservation of entropy, we have 
\bea\label{mag_entropy_conv}
a_0^4\frac{\pr \rho_B}{\pr \ln k}\bigg|_0 &=& a_f^4 \frac{\pr \rho_B}{\pr \ln k}\bigg|_{\eta_f}\nno\\
\Rightarrow \frac{\pr \rho_B}{\pr \ln k}\bigg|_0 &=&\left(\frac{a_f}{a_0}\right)^4 \frac{\pr \rho_B}{\pr \ln k}\bigg|_{\eta_f}.
\eea
Where ``0" denotes the present epoch, $a_0$ represents the scale factor at present, $\eta_f$ is the conformal time at the end of inflation, and $a_f$ is the scale factor corresponding to $\eta_f$. Implementing Eq. \eqref{mag_entropy_conv}, we can evaluate the present-day magnetic field strength.
\section{Calculation of total e-folding number of inflation}\label{appen3}
We have the expression of the total e-folding number of inflation $N_k$ from \cite{Maity:2021qps} as
\beq\label{inf_efold}
N_k =\int_{t_k}^{t_f}H(t)dt.
\eeq
The Hubble parameter explicitly depends on the background evolution. Therefore, to calculate the actual Hubble parameter, we will Taylor expand around the conformal time of horizon crossing $t_k$ to incorporate the background effects,
\beq\label{hubble_expand}
H(t) = H_k + \dot{H}_k (t-t_k) + \frac{1}{2} \ddot{H}_k (t-t_k)^2.
\eeq
In the above Eq. \eqref{hubble_expand}, we consider only terms up to ${\cal O}(\ddot{H}_k)$. The total duration of the inflation is represented as $\Delta t = t-t_k$. Then, by Eq. \eqref{hubble_expand}, the Hubble parameter at the end of inflation can be written as
\beq\label{hubble_end_inf}
H_f = H_k + \dot{H}_k \Delta t + \ddot{H}_k(\Delta t)^2. 
\eeq
Consequently, the duration of inflation $(\Delta t)$ can be expressed in terms of the Hubble parameter and the derivative of it as
\beq
\Delta t= \frac{|\dot{H}_k|}{\ddot{H}_k}\left(1-\sqrt{1-\frac{2 \ddot{H}_k}{|\dot{H}_k|^2}(H_k - H_{f})}\right).
\eeq
Finally, the total e-folding number during inflation turns out to be,
\bea\label{inf_efold1}
N_k &=&\int_{t_k}^{t_f}H(t)dt\nonumber\\
&=&\frac{H_k |\dot{H}_k|}{\ddot{H}_k}\left(1-\sqrt{1-\frac{2 \ddot{H}_k}{|\dot{H}_k|^2}(H_k - H_{f})}\right) 
- \frac{|\dot{H}_k|^3}{2 \ddot{H}_k^2}\left(1-\sqrt{1-\frac{2 \ddot{H}_k}{|\dot{H}_k|^2}(H_k - H_{f})}\right)^2\nonumber\\
&+&\frac{|\dot{H}_k|^3}{6 \ddot{H}_k^2}\left(1-\sqrt{1-\frac{2 \ddot{H}_k}{|\dot{H}_k|^2}(H_k - H_{f})}\right)^3 .
\eea
The slow-roll parameters are also connected through the Hubble parameter and its derivatives. In terms of the inflationary Hubble parameter, the scalar perturbation amplitude ${\cal A}_s$ and the scalar spectral index $n_s$ are related as
\bea\label{hd_hdd}
|\dot{H}_k| = \frac{H_k^4}{4\mathcal{A}_s M^2_{Pl}},\quad
\ddot{H}_k= \frac{H_k^5}{4\mathcal{A}_s M^2_{Pl}}\left(\frac{H_k^2}{\mathcal{A}_s M^2_{Pl}} - (1- n_s) \right).
\eea
The Eq.\eqref{inf_efold1} can also be inverted to take $N_k$ as an input parameter, and correspondingly, we can calculate the quantity $H_f$. For this study, we have taken  $N_k=50$ and get $H_f\sim 10^{13}$ GeV.
\section{Magnetic field power spectra during reheating era}
The power spectrum of the magnetic field in the post-inflationary era becomes
\bea
{\cal P}_B(\eta,\tk) &=& \frac{\tk^5}{\pi^2 a^4}\left(|u_1^{(re)}|^2+|u_2^{(re)}|^2+|u_3^{(re)}|^2\right)\
=\frac{\tk^4}{\pi^2 a^4}\sum_i  |\tu_i^{(re)}|^2\nno\\
&=& \frac{\tk^4}{\pi^2 a^4}\bigg[ \frac{1}{6\sqrt{3}}\bigg(1+2|\beta_1|^2+ 2\sqrt{1+|\beta_1|^2}|\beta_1|
\cos[Arg(\alpha_1\beta_1^*) -2\sqrt{3}\tk(\eta-\eta_f)]\bigg)\nno\\
&+& \frac{1}{24\sqrt{3}}\bigg(1+2|\beta_2|^2+ 2\sqrt{1+|\beta_2|^2}|\beta_2| 
\cos[Arg(\alpha_2\beta_2^*) -2\sqrt{3}\tk(\eta-\eta_f)]\bigg)\nno\\
&+& \frac{1}{24\sqrt{3}}\bigg(1+2|\beta_3|^2+ 2\sqrt{1+|\beta_3|^2}|\beta_3|
\cos[Arg(\alpha_3\beta_3^*) -2\sqrt{3}\tk(\eta-\eta_f)\bigg)\bigg],
\eea
where we have substituted the mode function solutions during the reheating era in terms of the Bogoliubov coefficients. The term $\eta-\eta_f$ is calculated as
\beq
\eta-\eta_f=\int \frac{da}{a^2 H}.
\eeq
Using the proper relations and substituting the value, we get the power spectrum of the magnetic field at the end of reheating as
\bea\label{pow_spec_reh1}
{\cal P}_B(\tk,\eta)\bigg|_{re}
&=&\frac{\tk^4}{\pi^2 a_{re}^4}\Bigg[\frac{1}{6\sqrt{3}}\bigg( 1+2|\beta_1|^2+ 2\sqrt{1+|\beta_1|^2}|\beta_1| \cos\left[Arg(\alpha_1\beta_1^*)
-\frac{4\sqrt{3}\tk}{(1+3 \omega_{eff})a_f H_f}\left(\left(\frac{H_f}{H_{re}}\right)^{\delta}-1\right)\right]\bigg)\nno\\
&+& \frac{1}{24\sqrt{3}}\bigg( 1+2|\beta_2|^2+ 2\sqrt{1+|\beta_2|^2}|\beta_2| \cos\left[Arg(\alpha_2\beta_2^*)
-\frac{4\sqrt{3}\tk}{(1+3 \omega_{eff})a_f H_f}\left(\left(\frac{H_f}{H_{re}}\right)^{\delta}-1\right)\right]\bigg)\nno\\
&+& \frac{1}{24\sqrt{3}}\bigg( 1+2|\beta_3|^2+ 2\sqrt{1+|\beta_3|^2}|\beta_3| \cos\left[Arg(\alpha_3\beta_3^*)
-\frac{4\sqrt{3}\tk}{(1+3 \omega_{eff})a_f H_f}\left(\left(\frac{H_f}{H_{re}}\right)^{\delta}-1\right)\right]\bigg)\Bigg].
\eea
\end{widetext}
\end{appendix}

\end{document}